\documentclass[prx,twocolumn, showpacs, floatfix,tightenlines amsmath, amssymb, preprintnumbers, superscriptaddress, longbibliography]{revtex4-1}

\usepackage{graphicx}
\usepackage{color}
\usepackage{dcolumn}
\usepackage{bm}
\usepackage{hyperref}
\usepackage[nolist,nohyperlinks]{acronym}
\usepackage{CJK}
\usepackage{braket}
\usepackage{nicefrac}
\usepackage{amsmath}
\definecolor{cblue}{RGB}{19,107,192}
\hypersetup{colorlinks=true,linkcolor=cblue,citecolor=cblue,urlcolor=cblue}

\newcommand{\GTS}{GaTa$_{4}$Se$_8$}
\newcommand{\GNS}{GaNb$_{4}$S$_8$}

\newcommand{\Ts}{T$^{*}$}

\begin{document}
\title{Bond Ordering and Molecular Spin-Orbital Fluctuations in the Cluster Mott Insulator \GTS{}}

\author{Tsung-Han Yang}
\affiliation{Department of Physics, Brown University, Providence, RI 02912, USA}

\author{S. Kawamoto}
\affiliation{Institute for Solid State Physics, University of Tokyo, Kashiwa, Chiba 277-8581, Japan}

\author{Tomoya Higo}
\affiliation{Department of Physics, The University of Tokyo, Tokyo  113-0033, Japan}

\author{SuYin Grass Wang }
\affiliation{NSF’s ChemMatCARS Beamline, The University of Chicago, Advanced Photon Source, Argonne, Illinois 60439, United States}

\author{M. B. Stone}
\author{Joerg Neuefeind}
\affiliation{Neutron Scattering Division, Oak Ridge National Laboratory, Oak
Ridge TN 37831-6573, USA}

\author{Jacob P. C. Ruff}
\affiliation{Cornell High Energy Synchrotron Source, Cornell University, Ithaca, NY 14853, USA}

\author{A. M. Milinda Abeykoon}
\affiliation{Photon Sciences Division, Brookhaven National Laboratory, Upton, New York 11973, USA}

\author{Yu-Sheng Chen}
\affiliation{NSF’s ChemMatCARS Beamline, The University of Chicago, Advanced Photon Source, Argonne, Illinois 60439, United States}

\author{S. Nakatsuji}
\affiliation{Institute for Solid State Physics, University of Tokyo, Kashiwa, Chiba 277-8581, Japan}
\affiliation{Department of Physics, The University of Tokyo, Tokyo  113-0033, Japan}
\affiliation{Institute for Quantum Matter and Department of Physics and
    Astronomy, The Johns Hopkins University, Baltimore, MD 21218, USA}
\affiliation{Trans-scale Quantum Science Institute, University of Tokyo, Tokyo  113-0033, Japan}

\author{K. W. Plumb}
\affiliation{Department of Physics, Brown University, Providence, RI 02912, USA}

\date{\today}
\begin{abstract}
  For materials where spin-orbit coupling is competitive with electronic correlations, the spatially anisotropic spin-orbital wavefunctions can stabilize degenerate states that lead to many and diverse quantum  phases of matter. Here, we find evidence for a dynamical spin-orbital state preceding a \Ts{}=50~K order-disorder spin-orbital ordering transition in the $j\!=\!3/2$ lacunar spinel \GTS{}.  Above \Ts{}, \GTS{} has an average cubic crystal structure, but total scattering measurements indicate local non-cubic distortions of Ta$_4$ tetrahedral clusters for all measured temperatures $2 < T < 300$~K.  Inelastic neutron scattering measurements  reveal the dynamic nature of these local distortions through symmetry forbidden optical phonon modes that modulate $j\!=\!3/2$ molecular orbital occupation as well as intercluster Ta-Se bonds. Spin-orbital ordering at \Ts{} cannot be attributed to a classic Jahn-Teller mechanism and  based on our findings, we propose that intercluster interactions acting on  the scale of \Ts{} act to break global symmetry. The resulting staggered intercluster dimerization pattern doubles the unit cell, reflecting a spin-orbital valence bond ground state.
\end{abstract}

\maketitle
\section{Introduction}
It is a maxim of condensed matter physics that the many and varied properties of strongly correlated materials arise from an intricate interplay of spin, orbital, and lattice degrees of freedom.
Often, the spin and orbital degrees of freedom have a mutual influence, but can be treated as distinct energy scales. Static orbital configurations impart a spatial anisotropy to affect many and diverse magnetic phenomena. For instance, orbital ordering can influence magnetic exchange interactions  to generate magnetic frustration in nominally unfrustrated lattices, \cite{Feiner:1997} or relieve magnetic frustration and drive valence bond solid transitions in frustrated magnets \cite{Pen:1997, DiMatteo:2004, Jackeli:2008}. Orbital overlap in metals can create one-dimensional bands in nominally three dimensional materials. The resulting Peierls instability on those bands generates intricate superstructures of spin singlets on structural dimers through an orbitally induced Peierls mechanism \cite{Radaelli:1997, Schmidt:2004, Khomskii:2005}.

\begin{figure}[t!]
    \includegraphics{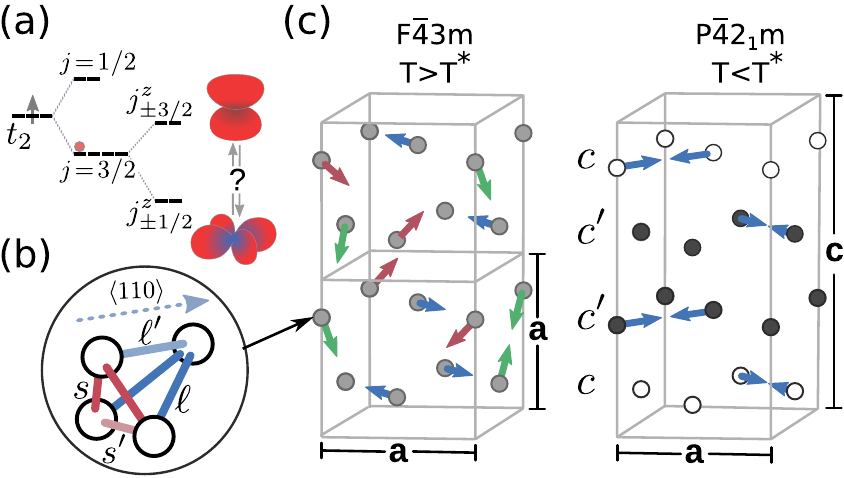}
\caption{Unit cell doubling and intercluster dimers in \GTS{}. (a) Spin-orbit coupling driven $j\!=\!3/2$ states on Ta$_{4}$ cluster molecular orbitals. $j^{z}$ spin-orbital states are Jahn-Teller degenerate. (b) Monoclinically distorted Ta$_{4}$ clusters with four inequivalent Ta-Ta distances $\ell\! >\! \ell^{\prime}\! >\! s^{\prime}\! >\! s $  point along local $\langle 110 \rangle$ axis. (c) Local distortions of Ta$_{4}$ clusters on FCC lattice  fluctuate among  $\langle 110 \rangle$ directions for T$>$\Ts{}, shown as colored arrows. Intercluster interactions select a global rotational symmetry breaking below \Ts{}. Unit cell doubling occurs through a formation of two inequivalent Ta$_{4}$ stacked in a $\!c\!-c^{\prime}\!-\!c^{\prime}\!-\!c$ pattern.}
\label{fig:schematic}
\end{figure}

\begin{figure*}[!t]
    \includegraphics{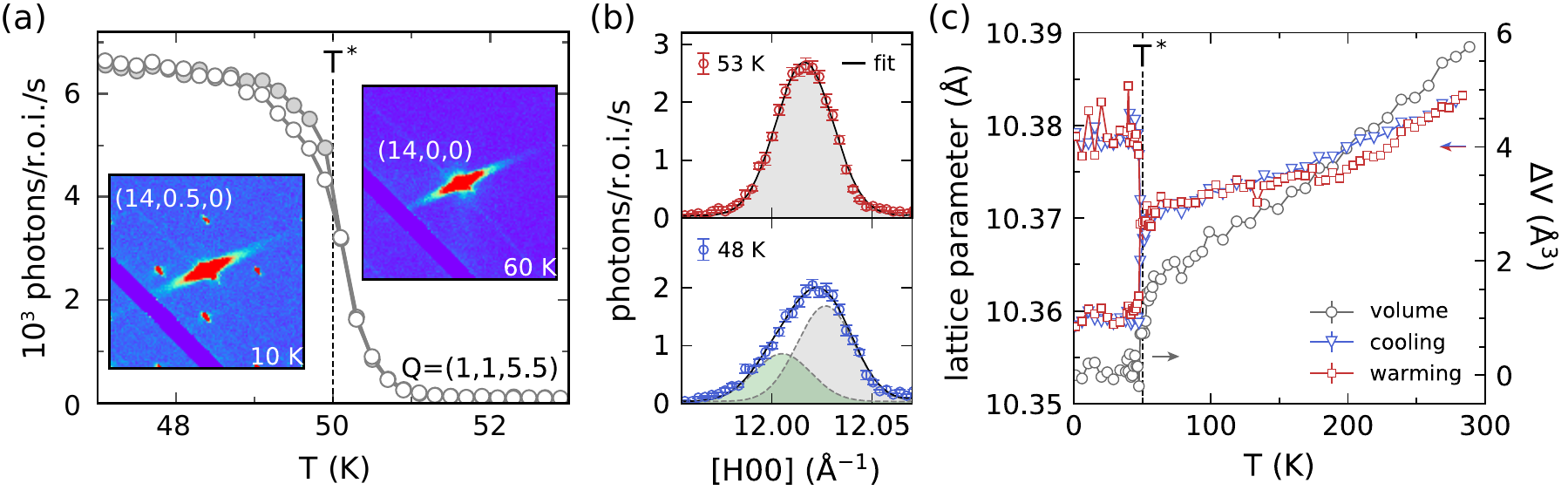}
\caption{ (a) X-ray diffraction reveals a first-order onset of superlattice $[1/2,0,0]$ at \Ts{}. (b) Bifrucation of $[12,0,0]$ reflection at \Ts{} indicating a low temperature tetragonal cell. (c) Temperature dependence of lattice parameter and cell volume extracted from the $[12,0,0]$ reflection. There is a sharp decrease in volume at \Ts{} but no measurable hysteresis.}
\label{fig:xrd}
\end{figure*}

Both relativistic spin-orbit coupling and covalency can act to reduce the separation of spin and orbital energy scales, enabling quantum fluctuations between nearly degenerate states. In the extreme limit of large spin-orbit coupling orbital degeneracy is either partially or totally removed depending on the electron filling. For instance, for a single electron occupying a $t_{2}$ orbital triplet, spin orbit coupling stabilizes a total $j\!=\!3/2$ quartet where the resulting spin-orbital wavefunctions alter Jahn-Teller potentials to promote fluctuations among many possible structural distortions [Fig.~\ref{fig:schematic} (a)] \cite{Streltsov:2020}. Such spin-orbital degrees of freedom naturally support spatially anisotropic and higher order, multi-polar exchange interactions \cite{Chen:2010, Jackeli:2009, Romhanyi:2017}. Covalency plays a similar, but lesser explored, role to reduce the tendency towards classical Jahn-Teller distortions in solid state materials \cite{Cianchi:1973, Whangbo:2002, Zeng:2011}. Typically, the effective strength of spin-orbit coupling is reduced in covalent materials, but, in materials where small clusters of atoms hybridize to form molecular units, the effects can be competitive. In such cluster Mott insulators, the interplay of orbitals, magnetism, and spin-orbit coupling on the molecular clusters leads to diverse possibilities \cite{Attfield:2015}.

In this work, we use x-ray diffraction, total scattering, and neutron spectroscopy to reveal dynamic spin-orbital fluctuations preceding a spin-orbital ordering transition in the $j\!=\!3/2$ cluster Mott insulator \GTS{}. At high temperatures, we find an average cubic crystal structure, but local cubic symmetry is  dynamically broken on Ta$_{4}$ clusters for all measured temperatures, T$\leq\!300$~K. The Ta$_{4}$ clusters distort  along local  $\langle 110 \rangle$ directions and fluctuate incoherently in space and time [Fig.~\ref{fig:schematic}]. Global symmetry is broken at \Ts{}=50~K as clusters coherently align towards a single neighboring cluster through a collective spin-orbital ordering transition. The distorted clusters exhibit an anti-ferro configuration doubling the cubic unit cell in a bond ordered structure as illustrated in Fig.~\ref{fig:schematic} (c).  Inelastic neutron scattering measurements directly resolve the  fluctuations associated with the disordered state above \Ts{} through the appearance of symmetry forbidden optical phonon intensity. These lattice fluctuations are a  manifestation of the dynamic spin-orbital state arising from a highly degenerate potential.  We propose that intercluster spin-orbital exchange interactions act to break the  degeneracy at \Ts{} and stabilize a long-range spin-orbital order in \GTS{}.

\GTS{} belongs to a family of cluster Mott insulating lacunar spinels with chemical formula GaM$_{4}$X$_{8}$, M = (V, Mo, Nb, Ta), X =(S, Se) \cite{Pocha:2000, Elmeguid:2004, Pocha:2005, Jakob:2007, Bichler:2008, Sieberer:2007, Camjayi:2012, Malik:2013}. The essential structural units are cubane (M$_{4}$X$_{4})^{5+}$ clusters that occupy an FCC lattice.  For M =(V, Nb, Ta), a single unpaired electron is localized on the cubane tetrahedral transition metal clusters. The transition metal atomic orbitals hybridize across $M_{4}$ clusters to  form molecular orbital wavefunctions with the highest occupied molecular orbitals on a tetrahedron comprised of triply degenerate $t$ states as shown in Fig.~\ref{fig:schematic} (a) \cite{Elmeguid:2004}.  As a consequence of the Jahn-Teller active molecular orbital, many of the lacunar spinels exhibit a typical separation of spin and orbital energy scales whereby a high temperature cooperative Jahn-Teller, orbital ordering, structural transition precedes a lower temperature magnetic ordering \cite{Pocha:2000, Pocha:2005, Jakob:2007}. However, in those compounds with heavy transition metals (Nb, Ta), such a separation of spin and orbital energy scales is either drastically diminished or does not exist \cite{Elmeguid:2004, Waki:2010, Ishikawa:2020}.
For both \GTS{} and \GNS{}, the spatial extent of transition metal wave functions and strong atomic spin-orbit coupling act to control the ground states and a single temperature scale \Ts{} defines magnetic and structural transitions \cite{Elmeguid:2004,Kawamoto:2016,  Geirhos:2021}.

\begin{figure*}[!t]
    \includegraphics{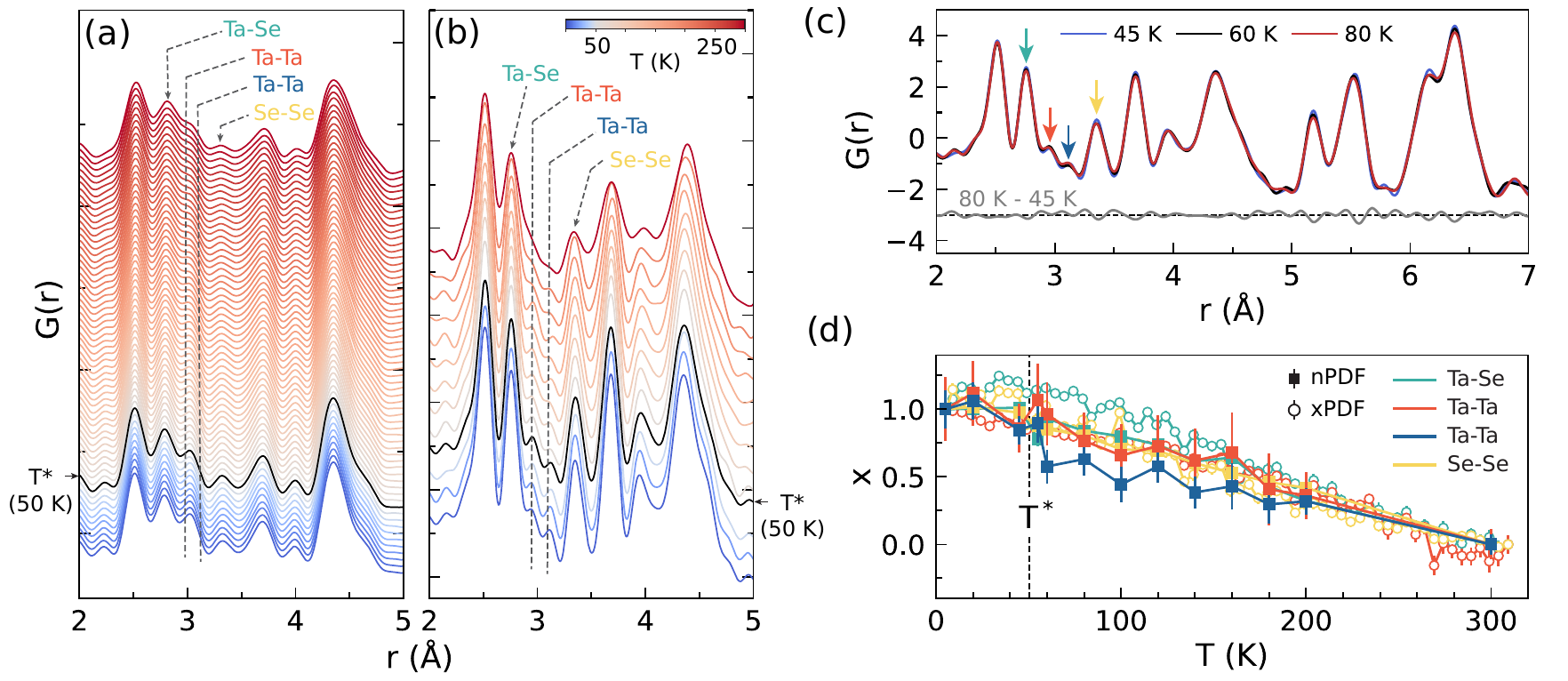}
\caption{Temperature dependence of atomic pair separation in \GTS{}. (a) Temperature dependent xPDF and (b) nPDF measurements show local distortions of Ta$_{4}$ clusters exist well above the transition through two distinct intercluster Ta-Ta distances, $s=$2.95~\AA~and $\ell=$3.06~\AA~observed at all measured temperatures. Real space resolutions of PDF data are  $\delta r\!=\!0.233$~\AA{} for xPDF and $\delta r\!=\!0.182$~\AA{} for nPDF  (c) Detailed comparison of small $r$ nPDF above and below T$^{*}$. 45 - 80 K difference data show no measurable change in atomic pair displacements through the transition. (d) Temperature dependence of normalized inter-Ta$_{4}$ cluster PDF peak intensities $x$ showing the continuous and smooth evolution of atomic distances with temperature. Filled symbols are from nPDF and open from xPDF.}
\label{fig:pdf1}
\end{figure*}

The importance of spin-orbit coupling in \GTS{} is borne out through density functional theory (DFT) and resonant inelastic x-ray scattering that suggest the molecular orbitals form a spin-orbit entangled $j=3/2$ quartet [Fig.~\ref{fig:schematic} (a)], analogous to the case of octahedrally coordinated d$^{1}$ Mott insulators \cite{KimHS:2014, Jeong:2017, Jeong:2021}. Such  $j=3/2$ spin-orbit entangled molecular orbitals are consistent with the reported small paramagnetic moment and entropy released at \Ts{} \cite{Kawamoto:2016, Ishikawa:2020}. In this case, the magnetic ground state below \Ts{} is formed from spin-orbital singlets as predicted for $j=3/2$ on an FCC lattice \cite{Romhanyi:2017, Jackeli:2009} with profound implications for the electronic properties of \GTS{}. Above \Ts{} and at ambient pressure, \GTS{} is a strongly correlated insulator, but it exhibits an insulator to metal and superconducting transition under the application of hydrostatic pressure \cite{Elmeguid:2004, TaPhuoc:2013, Camjayi:2014}.  Molecular $j=3/2$ degrees of freedom would support a topological superconducting state at high pressure \cite{Park:2020, Jeong:2021}. However, the $j=3/2$ quadruplet on $t_{2}$ orbitals is also Jahn-Teller active. For atomic $j=3/2$ insulators, spin-orbit coupling acts to suppress Jahn-Teller distortions  and, to first order, octahedral compression and elongation provide equal energy gain. This additional degeneracy results in a non-adiabatic ``Mexican Hat'' potential energy surface that promotes quantum fluctuations among different distortions \cite{Streltsov:2020}. Weak lattice anharmonicity or additional electronic interactions, may then act to ultimately select a distortion that is different from what a conventional Jahn-Teller effect would predict. A similar situation may apply to the molecular orbitals of \GTS{}; however, currently the low temperature crystal structure is unknown and there is little information on the spin-orbital configuration below \Ts{}.

There is indirect evidence for a magnetic singlet, valence bond solid like, ground state from neutron diffraction, $\mu$SR, and NMR \cite{Kawamoto:2016, Ishikawa:2020}. However, there is no information about the actual valence bond covering on the  highly frustrated on the FCC lattice of \GTS{} and limited  understanding of the intercluster interactions that might generate such spin-orbital singlets across molecular orbitals. Furthermore, expectations of strong Ta-Se covalency and configurational mixing not captured by DFT  bring the $j=3/2$ picture into question \cite{Majumdar:2004, Hozoi:2020, Petersen:2022}. Given the strong connection between the spin, orbital, and lattice degrees of freedom and the lattice,  a detailed account of the evolution of the lattice through \Ts{} is essential to understand the low temperature electronic state and it's connection to high pressure superconductivity in \GTS{}.

\section{X-ray crystallography and Unit Cell Doubling}
\GTS{} has a single magnetic and structural transition at \Ts{}\!=\!50~K where the magnetic susceptibility abruptly drops and the  unit cell doubles, as indicated by the sudden appearance of structural superlattice Bragg reflections at $[1/2,0,0]$ positions [Fig.~\ref{fig:xrd} (a)].  As shown in Fig.~\ref{fig:xrd} (b) and (c), a bifurcation of cubic $[h,0,0]$ reflections and abrupt volume decrease at \Ts{} indicate a first-order cubic to tetragonal transition, but with no detectable hysteresis or phase coexistence.  Above \Ts{}, single crystal XRD consistently refines to the cubic space group $F\bar{4}3m$. Below \Ts{}, the $[1/2,0,0]$ reflections indicate a unit cell doubling along a unique cubic axis and we observed all $[1/2,0,0]$ reflections consistent with six domains of the tetragonal cell. We did not find any additional peak splitting or evidence for an orthorhombic distortion. We have evaluated our T=10~K single crystal XRD data against all subgroups of the cubic cell that are consistent with a $k=[1/2,0,0]$ distortion; body centered space groups are ruled out by the observation of reflections with $h\!+\!k\!+\!l\!\neq\!2n$  and based on the absence of any $[2n\!+\!1,0,0]$ reflections we ruled out all possible primitive groups except for $P\bar{4}2_{1}m$ \cite{SI}.

We obtain high fidelity refinements for $P\bar{4}2_{1}m$ with $a=b=10.3437$ and $c=20.6878$ ($R_{1}=0.0879$ for 13104 reflections). Within the tetragonal cell there are two distinct Ta$_{4}$ clusters labeled $c$ and $c^{\prime}$ in Fig.~\ref{fig:schematic} (c), each having the same symmetry, but different Ta-Ta bond lengths. The symmetry of  tetrahedral clusters is reduced to monoclinic and retains a  single mirror plane. Each cluster contains four different bond lengths shown in Fig.~\ref{fig:schematic} (b) and listed in table ~\ref{tab:bonds}, with three long-bonds, labeled $\ell$ and $\ell^{\prime}$ ($\ell\!>\!\ell^{\prime}$), and three short-bonds labeled $s$ and $s^{\prime}$ ($s\!<\!s^{\prime}$) resulting in a characteristic elongation of tetrahedra along cubic \emph{face diagonals}. Each Ta$_{4}$ cluster points along $\langle 110 \rangle$ towards a single neighboring cluster, forming an intercluster dimer motif as illustrated [Fig.\ref{fig:schematic} (b) and (c)]. The cubic unit cell doubling occurs through a $c\!-\!c^{\prime}\!-\!c^{\prime}\!-\!c$ stacking arrangement of Ta clusters along the c-axis, giving rise to $[1/2,0,0]$ reflections from six domains of the tetragonal cell.

The structural transition in \GTS{} is distinct from reports in other lacunar spinels. For Ga(V,Nb)$_{4}$(S,Se)$_{8}$, the  Jahn-Teller active M$_{4}$ clusters  undergo a rhombahedral distortion, displacing along the cubic \emph{body diagonals}, in a ferrodistortive pattern for vanadium based compounds  \cite{Pocha:2000, Ruff:2015, Xu:2015, Wang:2015, Ruff:2017}  and antiferrodistortive pattern for the niobium compounds \cite{Jakob:2007, Geirhos:2021}. A primary mode analysis \GNS{} found a single  Jahn-Teller active mode within  irreducible representation $X_{5}$ and dominated by displacements of the Nb$_{4}$ tetrahedra \cite{Geirhos:2021}. In contrast, we find that the distortion in \GTS{} is described by a superposition of modes within many irreducible representations. The most dominant distortions belong to  $X_{5}$ (subgroup $P\bar{4}2_1m$, amplitude$=0.12$~\AA{}) and $X_{3}$ (subgroup $P\bar{4}m2$, amplitude$=0.07$~\AA) representations, both modes involving nearly equivalent displacements of Ta and Se atoms \cite{Orobengoa2009, PerezMato2010, SI}. Unit cell doubling is affected through a combination of $W_{4}$ and $\Delta_{1}$ modes with amplitudes of 0.02~\AA{} and nearly equivalent Ta/Se displacements \cite{SI}. Thus, we cannot attribute the transition in \GTS{} to a classic Jahn-Teller mechanism acting through a single normal mode of the cell.  The intercluster dimer motif, unit cell doubling,  mode mixing, and equivalent contributions from displacements of metal and ligand ions implicates the complex spin-orbital wavefunction and a mechanism likely involving intersite interactions, are at play in \GTS{}.

\section{Local Distortions and Dynamic Orbital Degeneracy Breaking}
\begin{figure*}[!t]
    \includegraphics{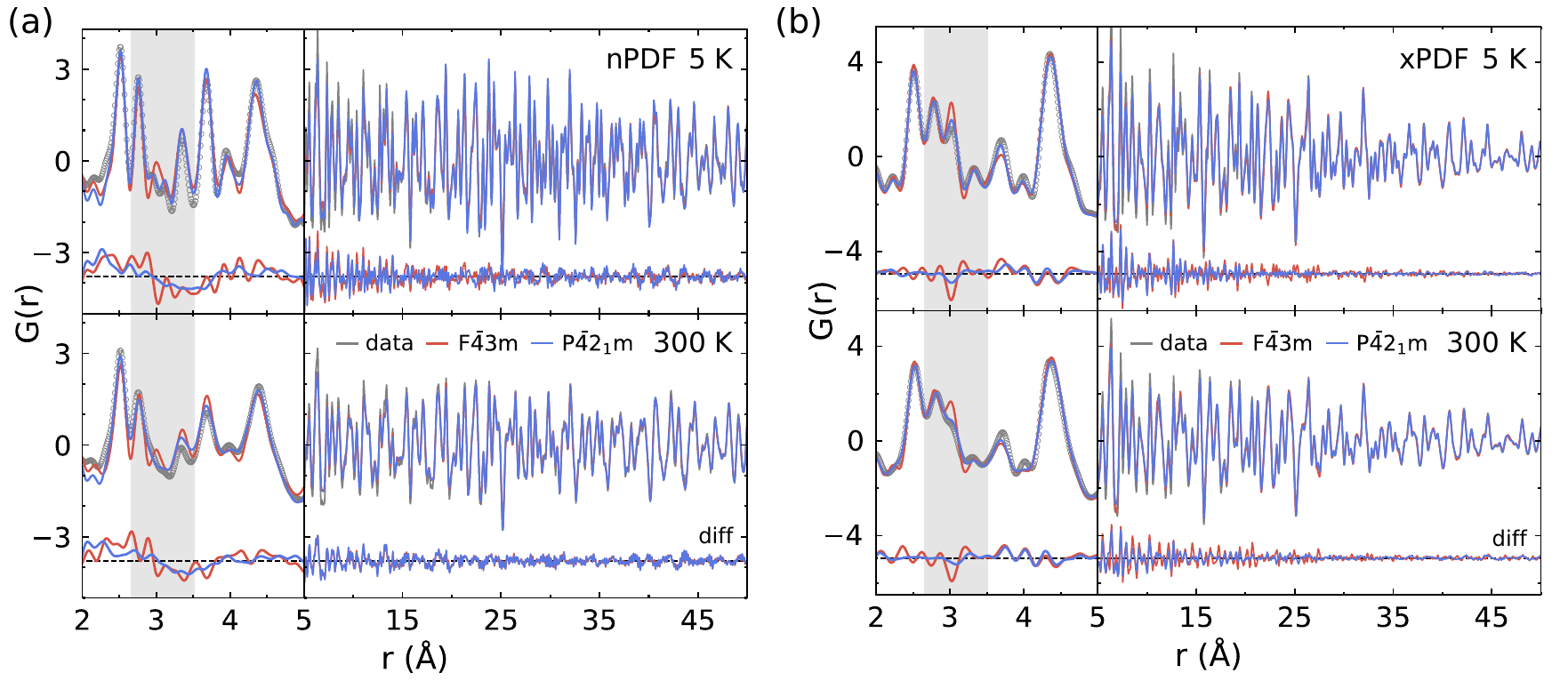}
\caption{Local and average structures of \GTS{}. Model fits to neutron pair distribution function measurements at  T = 5 K and T = 300 K for (a) nPDF and (b) xPDF. At all temperatures the tetragonal cell (P$\bar{4}2_{1}m$)  with distorted tetrahedra provides a significantly improved description of the local structure for $r\!<\!5$\AA, while the average structure is equally well described using the cubic or tetragonal cell. }
\label{fig:pdf2}
\end{figure*}

Having established the low temperature space group in \GTS{}, we now turn to atomic pair distribution analysis (PDF) to provide detailed insight into the evolution of atomic correlations across different length-scales through \Ts{}. Temperature dependent x-ray (xPDF) and neutron (nPDF) PDF measurements shown in Fig.~\ref{fig:pdf1} provide a histogram of  atomic pair correlations from which atomic pair separations can be read off independent of a specific structural model. The PDF data for \GTS{} have two striking features. First, the double-peak structure around 3~\AA{} indicates at least two intercluster Ta-Ta distances between 2.9 and 3.1~\AA{} for temperatures up to 300~K. Thus, the local crystal structure of \GTS{} is not cubic well into the temperature regime where the average structure is cubic. Second, nPDF and xPDF data do not show any discernible qualitative changes across all measured temperatures beyond thermal broadening. The absence of any change in the local structure is highlighted in Fig.~\ref{fig:pdf1} (c) that shows a negligible difference in  nPDF across \Ts{} and in Fig.~\ref{fig:pdf1} (d) that shows the normalized inter-Ta$_{4}$ cluster PDF peak intensities
\begin{equation}
x(r^{\prime},T) = \frac{G(r^{\prime},T) - G(r^{\prime},300)}{G(r^{\prime}, 5) - G(r^{\prime},300)},
\end{equation}
for inter and intra-cluster pair separations.  $x$ acts as an effective local order parameter for displacive transitions \cite{Thygesen:2017}. The smooth evolution of $x$ with temperature for all pair separations is consistent with thermal disorder (Debye-Waller) and directly implicates an order-disorder transition at \Ts{}. Local distortions in \GTS{} are present but uncorrelated across length scales beyond one unit cell at least up to 300~K. At \Ts{}, global cubic symmetry is broken as the distorted units select a unique $\langle 110 \rangle$ axis and stack in a  $\!c\!-c^{\prime}\!-\!c^{\prime}\!-\!c$, forming the cell doubled tetragonal structure of Fig.~\ref{fig:schematic} (b).

In Fig.~\ref{fig:pdf2}, we show a model dependent analysis of the PDF data, covering  broad length and temperature scales. A cubic model fails to account for the observed local structure ($r\!<5$~\AA) for all measured temperatures T$<$~300~K. In particular, the cubic structure cannot capture the two Ta-Ta distance around 3~\AA{} and overestimates the PDF weight for Se-Se pairs at 3.5 \AA{}. This local structure is reproduced at all measured temperatures by the tetragonal $P\bar{4}2_{1}m$ lattice that includes four intercluster Ta-Ta bond lengths $\ell \! > \! \ell^{\prime} \! > \! s^{\prime} \! > \!s$. Within the accuracy of our measurements, the degree of local distortion in \GTS{} as measured by the bond length difference $\ell - s$ does not depend on temperature, indicating local distortions of Ta clusters are fully set in up to at least T=300~K \cite{SI}. On the other hand, the PDF data for $r\!>\!5$~\AA{} are equally well described by the cubic and tetragonal models at all temperatures because the tetragonal cell is metrically cubic ($c/2\!\sim\!a$). We find that  within experimentally achievable resolution, the average structures of  cubic and tetragonal cells are only distinguished in a PDF measurement through a small anomalous upturn in thermal parameters upon cooling through \Ts{} \cite{SI}.

Ta$_{4}$ tetrahedral clusters in \GTS{} are distorted at all measured temperatures, indicating that local, and potentially dynamic, Jahn-Teller distortions are already present at 300~K. Local distortions preceding an orbital ordering structural transition are common in transition metal compounds, but global symmetry breaking is generally also apparent in total scattering measurements through the appearance of a new length-scale or an abrupt change in bond lengths as atoms displace through the transition \cite{Radaelli:1997, Schmidt:2004,  Qiu:2005, Yan:2011, Thygesen:2017, Yan:2019, Koch:2021, Yang:2020}. The structural transition in \GTS{} is unique in this respect. An absence of any detectable redistribution in PDF intensity though \Ts{} demonstrates that no length-scale, or further local symmetry breaking, is introduced at the structural transition. Instead, a global six-fold discrete rotational symmetry breaking occurs at \Ts{}, where already distorted Ta$_{4}$ clusters select a particular $\langle 110 \rangle $ axis [Fig.~\ref{fig:schematic} (b)]. Superlattice reflections in a diffraction measurement are a consequence of this global rotational symmetry breaking.

\section{Phonon Density of States and Dynamic Distortions}
\begin{figure}[!t]
    \includegraphics{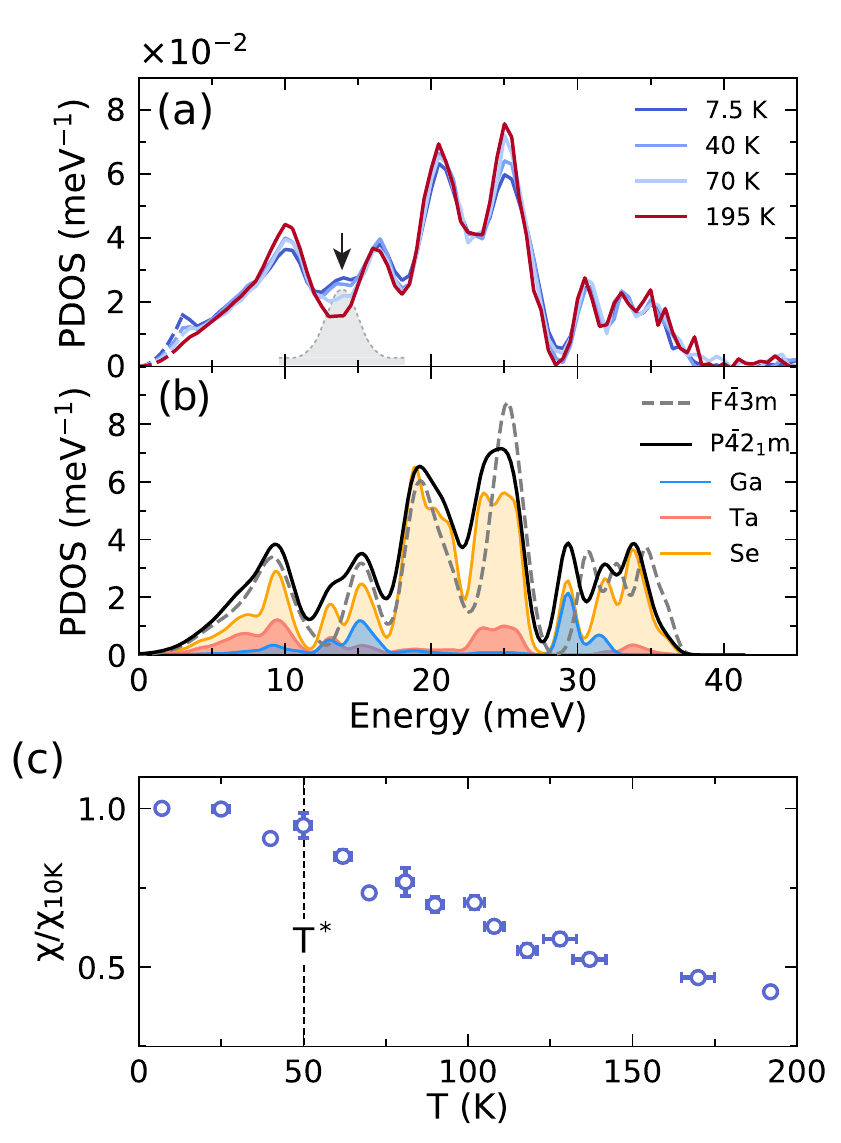}
\caption{Dynamic symmetry breaking in \GTS{}. (a) Measured neutron weighted phonon density of states in \GTS{}. A representative Lorentzian fit to the cubic symmetry forbidden optical phonon intensity at 13.8~meV is shown as the grey shaded area. (b) DFT calculated neutron weighted phonon DOS for the cubic (F$\bar{4}3$m) and tetragonal (P$\bar{4}2_{1}$m) cells. Partial contributions to the tetragonal cell DOS are also shown. (c) Temperature dependence of the 13.8 meV intensity extracted from Lorentzian fits to the dynamic susceptibility.}
\label{fig:4}
\end{figure}

The energy integrated diffraction measurements presented above find an average cubic structure for \GTS{} at T$>$\Ts{}, implying that local distortions are incoherently fluctuating in space, time, or both. In Fig.~\ref{fig:4} we show the measured phonon density of states (PDOS)  as a function of temperature that provide direct evidence for the dynamic nature of local distortions above \Ts{} through the appearance of a symmetry forbidden optical phonon mode (13~meV) well into the cubic phase.

Measured PDOS at select temperatures are shown in Fig.~\ref{fig:4} (a). The most striking feature of this data is a temperature dependent 13~meV phonon intensity appearing above background below T=200~K. This mode is visible in the raw neutron scattering data as a non-dispersive feature with an intensity that increases quadratically with momentum transfer, ruling out a magnetic origin \cite{SI}. In Fig.~\ref{fig:4} (c) we show the temperature dependent integrated intensity of the 13~meV mode, corrected for the Bose factor, and normalized to it's intensity at T=10~K. Above 200~K it was not possible to reliably resolve the 13~meV intensity from the adjacent modes at 10 and 15 meV. We find that the integrated intensity increases linearly with decreasing temperature from 200~K and saturates at \Ts{} while the energy position and width are constant and set by the energy resolution of the spectrometer \cite{SI}.

We have carried out first principles calculations of the neutron weighted PDOS for \GTS{} in order to obtain insight into the origin of the 13~meV intensity. The results of these calculations for both F$\bar{4}3m$ and P$\bar{4}2_{1}$m space groups are shown in Fig.~\ref{fig:4} (b) and are in excellent agreement with the measured spectra. There is a notable absence of phonon modes around 13~meV for the cubic cell; but, significant 13~meV PDOS weight arising from optical phonons in the tetragonal cell. Indeed, the 13~meV phonon intensity is the primary qualitative distinction between the cubic and tetragonal PDOS. The appearance of this optical phonon intensity for T$>$\Ts{}, where the average structure is cubic, directly reflects the dynamic structural fluctuations associated with the local symmetry breaking for T$>$\Ts{} in \GTS{} as schematically illustrated in Fig.~\ref{fig:schematic} (b). We have evaluated the partial contributions of Ga, Ta, and Se to the PDOS of the tetragonal cell and find the dominant contribution is from Se with additional secondary contributions from Ta and Ga [Fig.~\ref{fig:4} (b)].  Optical $A_{1}$, $A_{2}$, $B_{1}$, $B_{2}$, and $E$  phonon modes all contribute the PDOS between 12 to 14~meV and our data are not consistent with the assignment of a unique Jahn-Teller phonon mode on Ta$_{4}$ clusters to this feature of the PDOS.

It is instructive to compare the temperature dependent phonon spectra in \GTS{} measured by INS with that measured by infrared spectroscopy (IR). The PDOS as measured by INS contains the full spectrum of predicted vibrational modes, while IR measurements only find a subset of the symmetry allowed modes all attributed to vibrations on the ligand sites \cite{Reschke:2020}. Such a discrepancy must arise as a result of a substantial screening of vibrational modes from electron delocalization over Ta$_{4}$ tetrahedral units in insulating \GTS{}. Neutrons interact directly with nuclear cores and are not subject to electronic screening. Such an electronic delocalization is most apparent in \GTS{} because of the spatially extended nature of Ta d orbitals. The apparently metallic Ta$_{4}$ clusters are at odds with local Jahn-Teller mechanisms generally accepted as the origin of structural transitions in Lacunar spinels and evidence that a picture from mean field arguments for M$_{4}$ molecular clusters should be revisited, especially in the case of Ta and Nb compounds \cite{Majumdar:2004, Hozoi:2020, Petersen:2022}. For \GTS{}, a minimal description should involve active and hybridized Ta and Se orbitals on cubane clusters, spin-orbit coupling on the molecular orbitals, as well as interactions between clusters. The mixed character of dynamic distortions and decomposition of the structural transition into a superposition of irreducible representations  implicate a  degeneracy among possible Jahn-Teller distortions in \GTS{}. This is likely a consequence of spin-orbit coupling and covalency on the Ta$_{4}$ clusters. Ultimately, additional interactions  between clusters acting on the scale of \Ts{} select a unique valence bond state and quench the structural fluctuations.

\section{Discussion}
We find that local and dynamic distortions are present up to at least 300~K in \GTS{}. These reduce tetrahedral symmetry of the Ta$_{4}$ clusters and dynamically break orbital degeneracy. Heat capacity measurements have found an entropy change associated with \Ts{} approaching $\ln(4)$ \cite{Kawamoto:2016, Ishikawa:2020} indicating that the orbital and spin degrees of freedom simultaneously freeze out at \Ts{}. In other words, \Ts{} corresponds to a collective, order-disorder, spin-orbital ordering between two insulating phases.  A discrete six-fold rotational symmetry on the FCC lattice is broken through \Ts{} and we propose this order is selected through intercluster spin-orbital interactions that act on the scale of \Ts{}.


Interestingly, diffraction measurements indirectly implicate dynamic structural fluctuations above global structural transitions in other lacunar spinels \cite{Wang:2015, Waki:2010, Geirhos:2021}. However, the nature of structural and magnetic transitions, local distortions of transition metal clusters, low temperature crystal structure, and magnetic state are unique in \GTS{} among the family of d$^{1}$ Lacunar spinels. GaV$_{4}$(S,Se)$_{8}$ and GaNb$_{4}$Se$_{8}$ exhibit a typical separation of spin and orbital energy scales with rhombahedral structural distortions that quench orbital moments occurring at high temperatures preceding lower temperature spin only magnetic transitions \cite{Pocha:2000, Pocha:2005,Malik:2013, Wang:2015, Dally:2020, Ishikawa:2020}. GaNb$_{4}$S$_{8}$ exhibits a unit cell doubling spin-orbital transition  that is similar to \GTS{}, but the symmetry of the low temperature structure is further reduced from P$\bar{4}2_{1}m$ to P$2_{1}2_{1}2_{1}$ with Nb$_{4}$ clusters distorting along cubic body diagonal directions in an antiferro pattern \cite{Isobe:2002, Jakob:2007, Geirhos:2021}. The material trend implicates both spatially extended electron density and spin orbit coupling as these both increase from V$\rightarrow$Nb$\rightarrow$Ta and can respectively act to drastically alter Jahn-Teller physics  \cite{Streltsov:2020, Khomskii:2021}. In particular, for $j\!=\!3/2$ degrees of freedom, spin orbit coupling can lead to a degeneracy among different types of possible Jahn-Teller distortions \cite{Streltsov:2020}. The symmetry forbidden phonon intensity we observe above \Ts{} in \GTS{} is consistent with such a  degeneracy that is ultimately broken by additional interactions.

Above \Ts{}, \GTS{} is a correlated Mott insulator \cite{TaPhuoc:2013, Camjayi:2014}, containing an odd number of electrons in the cubic cell. While the Tetragonal unit cell below \Ts{} contains four formula units and an even number of valence electrons, our band structure calculations find a metallic state for this structure, supporting the Mott insulating picture. The absence of magnetic ordering must then indicate a valence bond solid type ground state. The structural dimerization suggests a staggered valence bond covering shown in Fig.~\ref{fig:schematic} (b).  Such a sharp reduction in magnetic susceptibility and dimerized superstructure below \Ts{} is remarkably similar to the orbitally induced Peierls metal-insulator transitions observed in spinels \cite{Radaelli:2002, Schmidt:2004,  Khomskii:2005} and other orbitally degenerate compounds exhibiting temperature dependent metal insulator transitions \cite{Okamoto:2008, Katayama:2009}. However, in \GTS{} the transition is between two insulating states.

Resonant inelastic x-ray scattering found that at the level of DFT the molecular orbital state in \GTS{} is molecular analog of the atomic $j=3/2$ \cite{Jeong:2017}.  Although, covalency, configurational interactions within cubane clusters, and local distortions alter the pure $j\!=\!3/2$ description \cite{Hozoi:2020, Petersen:2022}, it may still stand as an approximate description of the magnetism. In this case \GTS{} may realize a spin-orbital dimer ground state proposed for Mott insulators on an FCC lattice with d$^{1}$ filling and large spin-orbit coupling  \cite{Romhanyi:2017}. Our measurement show how the lattice degrees of freedom can act to select an ordered valence bond solid from the glassy spin-orbit dimer covering proposed in \cite{Romhanyi:2017}.

Our  measurements implicate an essential role played by cluster orientational ordering to affect a Peierls like intercluster dimerization at \Ts{}. This ordering is likely driven by  intercluster spin-orbit interactions that break the degeneracy of orbital configurations. The importance of such interactions is further underscored by the primary involvement of both Ta and Se ions across intercluster bonds through the structural transition. Despite the large intercluster separation, closest Ta-Ta distance 4.274~\AA{} closest Ta-Se distance 2.744~\AA{}, there is strong evidence that intercluster exchange is a relevant energy scale in lacunar spinels. For example, recent inelastic neutron scattering measurements of spin waves in magnetically ordered GaV$_{4}$S$_{8}$ suggest exchange interactions between the clusters on the order of 20~K \cite{Pokharel:2021}. A more spatially extended electron density on Ta 5d orbitals should give rise to enhanced interactions compared with the V compounds so that exchange interactions on the scale of \Ts{}$\!=\!50$~K are not unreasonable. In \GTS{} the broken symmetry at \Ts{} involves both spin and orbital degrees of freedom and such spin-orbital exchange may be mediated through lattice vibrations \cite{Goodenough:1961, Chibotaru:1988}. The 13~meV mode we observe distorts Ta$_{4}$ clusters, simultaneously modulating orbital occupancy, intercluster distances, and Ta-Se bond angles to influence intercluster interactions. In this respect, the optical phonon response we observe in \GTS{} is reminiscent of spin Peierls compound CuGeO$_{3}$ \cite{Hase:1993, Braden:1998}, but in the case of \GTS{}, lattice vibrations affect a complex spin-orbital exchange. Further theoretical effort investigating the role of optical phonons in generating effective spin-orbital interactions in cluster Mott insulators is required \cite{Zhou:2001, Chibotaru:2005}. Our results provide empirical insight on the microscopic mechanisms that may generate such orbital interactions.

In summary, we have used a combination of single crystal diffraction, total scattering, and inelastic neutron scattering measurements to show that Ta$_{4}$  tetrahedral clusters in \GTS{} are locally and dynamically distorted up to temperatures of at least 300~K. We resolve the optical phonon modes associated with lattice fluctuations between degenerate orbital configurations on the FCC lattice. We propose that the degeneracy of these configurations is broken through inter-cluster spin-orbit exchange interactions. A collective order-disorder spin-orbital transition in \GTS{} is characterized by the long range orientational ordering of distorted Ta$_{4}$ clusters to form weak structural dimers aligned along $\langle 110 \rangle$ directions. This suggests that a  spin-orbital valence bond solid ground state is realized in \GTS{}.


\begin{acknowledgments}
We are thankful to Arun Paramekanti, Ben Frandsen, Danilo Puggioni and Emil Bozin for helpful and informative discussions. Work at Brown University was supported by the U.S. Department of Energy, Office of Science, Office of Basic Energy Sciences, under Award Number DE-SC0021223. The work at the Institute for Quantum Matter, an Energy Frontier Research Center was funded by DOE, Office of Science, Basic Energy Sciences under Award DE-SC0019331. This work was partially supported by JST-CREST (JPMJCR18T3), and Grants-in-Aid for Scientific Research (19H00650).   A portion of this research used resources at the  Spallation Neutron Source, a DOE Office of Science User Facility operated by the Oak Ridge National Laboratory. This work used beamline 28-ID-1 beamline of the National Synchrotron Light Source II, a U.S. Department of Energy (DOE) Office of Science User Facility operated for the DOE Office of Science by Brookhaven National Laboratory under Contract No. DE-SC0012704.  NSF’s ChemMatCARS Sector 15 is supported by the Divisions of Chemistry (CHE) and Materials Research (DMR), National Science Foundation, under grant number NSF/CHE- 1834750. Use of the Advanced Photon Source, an Office of Science User Facility operated for the U.S. Department of Energy (DOE) Office of Science by Argonne National Laboratory, was supported by the U.S. DOE under Contract No. DE-AC02-06CH11357. This work is based upon research conducted at the Center for High Energy X-ray Sciences (CHEXS), which is supported by the National Science Foundation under award DMR-1829070.
\end{acknowledgments}

\appendix
\setcounter{equation}{0}
\renewcommand{\theequation}{A\arabic{equation}}
\section{Methods}

\subsection{Sample Synthesis}
Single crystal samples of \GTS{} were synthesized by solid-state reaction in box furnaces. About 1.5~g of stoichiometric quantities of Ga, Ta, and Se were loaded into a I.D.~=10~mm, 170~mm long quartz ampules under an argon glovebox, evacuated, and sealed. The ampules were heated to 1050~$^{\circ}$C with 80~$^{\circ}$C/hour ramping rate and stayed at 1050~$^{\circ}$C for 24 hours. Then cool down to 850~$^{\circ}$C with 1~$^{\circ}$C/hour cooling rate before air quench. We were able to obtain thousands of high-quality single crystals with a dimension above 200 microns in a single batch. The quality of single crystals was confirmed by x-ray diffraction (Ag K$\alpha$ radiation). The composition was confirmed by powder x-ray diffraction with ground crystals (Bruker Cu K$\alpha$ radiation).

\subsection{Synchrotron x-ray crystallography}
Temperature-dependent synchrotron x-ray diffraction measurements were carried out on the QM2 beamline at Cornell High Energy Synchrotron source using an x-ray energy of 20.67~keV (0.5998~\AA), and the Pilatus 300K area detector positioned 56.2 cm from the sample. The single crystal sample was mounted on a Cu post on the coldfinger of a closed cycle cryostat for measurements in reflection geometry between 3 and 300~K.

Synchrotron x-ray crystallography measurements were carried out on 15-ID at Advanced Photon Source, Argonne National Laboratory using an x-ray energy of 30.00~keV (0.4133~\AA) and Pilatus CdTe detector positioned at 7 cm from the sample. The ~20~$\mu$m single crystal was mounted on a quartz capillary, and measurements were carried out between 10 K and 300 K using a He cryostream. Diffraction peaks were indexed, unit cell, and UB matrix of each tetragonal domain were determined in reciprocal lattice and \textsf{CELLNOW} software for six domains. All frames were integrated using Bruker \textsf{SAINT} software, and multidomain absorption corrections were performed through \textsf{TWINABS}. The crystal structure was solved by direct method and processed least-square method refinements on $F^{2}$ by \textsf{SHELXTL} \cite{SHELX} and \textsf{Olex2} \cite{olex2} packages. The converged solution using full single domain reflection data and all Ga, Ta and Se atoms were modeled by isotropic atomic displacement parameters (ADPs). We checked $1\!\times\!1\! \times\!2$, $2\!\times\! 2\! \times\! 1$ and $2\! \times\! 2\! \times\! 2$ superlattices in $-4$ and $-42m$ tetragonal lattice classes. Based on the extinction rules of x-ray reflection, we find the P$\bar{4}$2$_{1}m$ space group with $1\!\times\!1\! \times\!2$ superlattice is the best structure for \GTS{} below the transition temperature. The details of the crystallographic analysis are shown in the supplementary material \cite{SI}.

\subsection{Pair distribution function measurements}
Synchrotron total scattering measurements (xPDF) were carried out on 28-ID-1 (PDF) beamline at the National Synchrotron Light Source-II at Brookhaven National Laboratory. The sample was loaded in a 1 mm I.D. kapton capillary, mounted in a He cryostat, and data was collected between 5~K and 330~K on warming in 5 K steps. Measurements were carried out using an x-ray energy of 74.46 keV (0.1665~\AA). We used an amorphous silicon PerkinElmer area detector with a sample to detector distance of 1008 mm for diffraction and 204 mm for PDF. Detector position was determined by calibration using a Ni standard. The two dimensional diffraction data were integrated using \textsf{PyFAI} \cite{ashiotis2015fast}  and then corrected, normalized, and Fourier transformed to obtain the atomic pair distribution function (PDF), $G(r)$, using \textsf{PDFgetX3} \cite{pdfgetx3} and a Q$_{max}$ = 27~\AA$^{-1}$.

Neutron total scattering measurements (nPDF) were carried out on the NOMAD beamline at Spallation Neutron Source at Oak Ridge National Laboratory. A 4.2~g sample was loaded into a vandadium sample can and mounted on the coldfinger of a He cryostat. Data was collected for 160~minutes at 5~K, 55~K and 300~K, and 80~minutes for intermediate temperatures. The structure factor S(Q) was checked to determine the Q$_{max}$ for each temperature. Background subtracted time-of-flight data were reduced and transformed to PDF using a Q$_{max}$ = 34.5~\AA$^{-1}$ and \textsf{ADDIE} available at NOMAD \cite{mcdonnell2017addie}. xPDF and nPDF refinements were carried out using \textsf{PDFgui} and \textsf{DIFFPY-CMI} modeling platforms \cite{Farrow:2007}.

\subsection{Inelastic neutron scattering}
Inelastic neutron scattering measurements were carried out on the SEQUIOA spectrometer at the Spallation Neutron Source at Oak Ridge National Laboratory. A 5 g polycrystalline sample was loaded into an Al sample can and mounted on the coldfinger of a closed cycle cryostat to achieve a base temperature of 5~K. Measurements were conducted using a fixed incident energy E$_{i}$ = 60 meV and the coarse Fermi chopper rotating at 300~Hz, providing an energy resolution at the elastic line of 3.1~meV. All data was normalized and reduced using algorithms in the \textsf{MANTID} analysis software \cite{Arnold:2014}. The normalized neutron intensities were converted into neutron weighted phonon density of states by integrating over Q (0.2 to 5 \AA$^{-1}$) and correcting for multiphonon and multiscattering processes within the incoherent approximation \cite{Lin:2018}.


\section{Density functional theory calculations}
\subsection{Electronic structure}
\begin{figure}[h!]
  \includegraphics{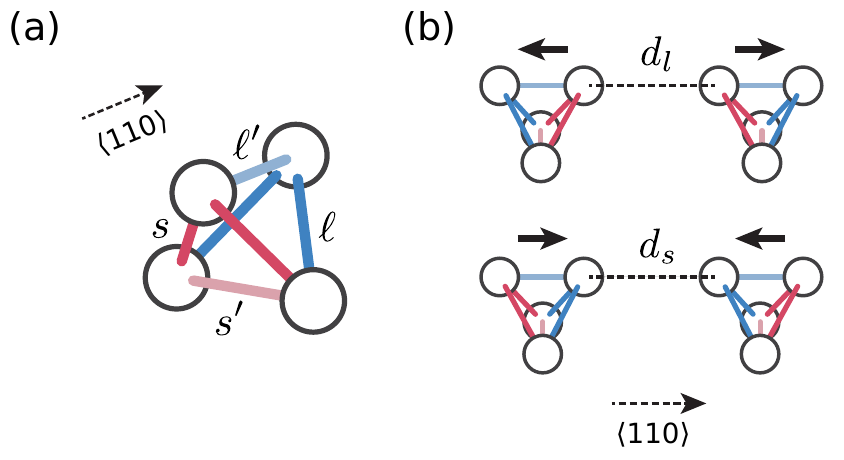}
\caption{(a) Local distortion of Ta$_{4}$ clusters, $s < s^{\prime} < l^{\prime} < l$. (b) The inter-cluster distances. The bond lengths for the two inequivalent small clusters $c$ and $c^{\prime}$ are shown in Table \ref{tab:bonds}.}
\label{fig:SITa4Clusters}
\end{figure}
We carried out density functional theory (DFT) calculations of the electronic structure of P$\bar{4}$2$_{1}$m \GTS{} without (PAW-PBE) and with spin-orbit coupling (PBE+SOC). All atomic positions and cell parameters were restricted by P$\bar{4}$2$_{1}$m space group and the force threshold was set to 4$\times$10$^{-5}$ a.u.. We used a 680 eV (PAW-PBE) and 620 eV (PBE+SOC) cutoffs for plane-wave energy with 6$\times$6$\times$6 (cubic) and  5$\times$5$\times$3 (tetragonal) Monkhorst-Pack k-point mesh for self-consistent-field calculations. The projector-augmented-wave (PAW) method \cite{PAW1994} and Perdew-Burke-Ernzerhof (PBE) exchange-correlation functional \cite{PBE1996} were used for all calculations. All calculations were performed using \textsf{Quantum ESPRESSO} \cite{giannozzi2009quantum,giannozzi2017advanced}.

The optimized structures are consistent with Ta$_{4}$ cluster distortions and intercluster dimerizations along $\langle$110$\rangle$ as we observed in x-ray diffraction. Details of the local Ta$_{4}$ cluster structures are presented in Fig. \ref{fig:SITa4Clusters} and the inter and intracluster bond lengths for PAW-PBE (PBE), PBE+SOC (SOC), and experimentally determined values are presented in table~\ref{tab:bonds}. The optimized structures display a slightly elongated c-axis (c$>$2a), which also agrees with our experimental result. In addition, we find that PAW-PBE overestimated the unit cell size (a$=$10.4677~\AA, c$=$21.0086 ~\AA) as PBE+SOC (a$=$10.3988~\AA, c$=$20.8129 ~\AA) has smaller cell and slightly closer to experimentally determined values (a$=$10.3437~\AA, c$=$20.6878~\AA). Also, Ta$_{4}$ clusters form dimers along $\langle$110$\rangle$ directions in both calculated structures.
\begin{table}
\caption{\label{tab:bonds}Intra and intercluster distances in~\AA. The definition of bonds are shown in Fig.~\ref{fig:SITa4Clusters}. $c$ and $c^{\prime}$ denote the inequivalent clusters in P$\bar{4}$2$_{1}$m tetragonal structure. PBE (PAW-PBE), SOC (PBE+SOC) are optimized structures from DFT.} \begin{ruledtabular}
\begin{tabular}{ c c c c c c c}

  & PBE ($c$) & PBE ($c'$) & SOC ($c$) & SOC ($c'$) & Exp ($c$) & Exp ($c'$) \\
$l$  & 3.129 & 3.129 & 3.084 & 3.084 & 3.06 & 3.04 \\
$l'$ & 3.086 & 3.086 & 3.053 & 3.053 & 3.03 & 3.03 \\
$s'$ & 2.939 & 2.939 & 2.951 & 2.951 & 2.97 & 2.97 \\
$s$  & 2.927 & 2.927 & 2.935 & 2.935 & 2.95 & 2.95 \\
$d_{l}$ & 4.358 & 4.357 & 4.331 & 4.332 & 4.30 & 4.29 \\
$d_{s}$ & 4.283 & 4.282 & 4.273 & 4.273 & 4.28 & 4.27



\end{tabular}
\end{ruledtabular}
\end{table}

\begin{figure}[t!]
  \includegraphics{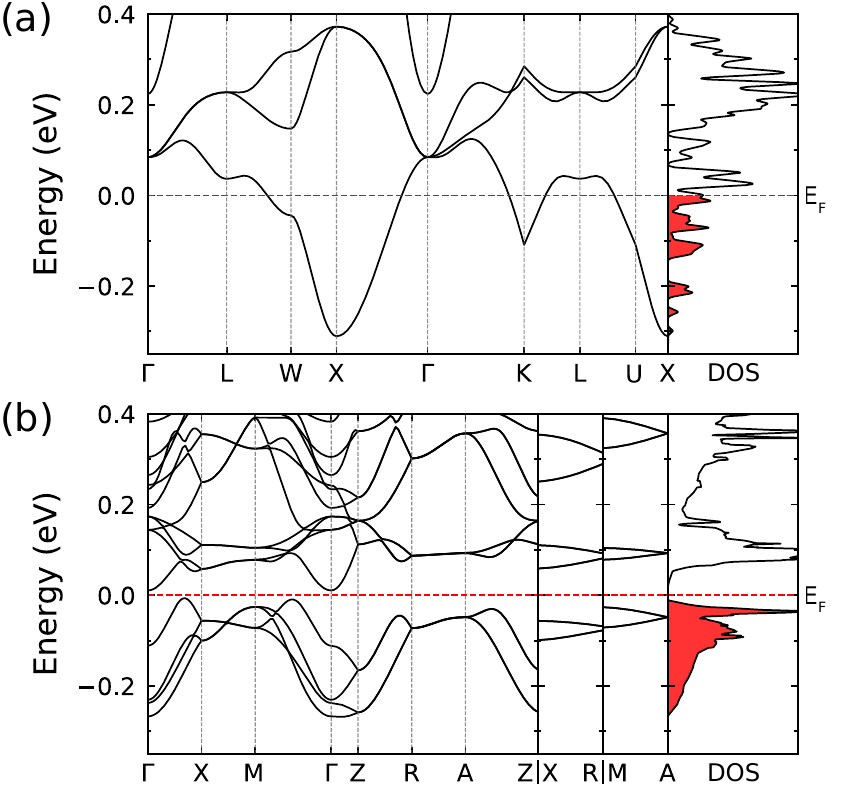}
\caption{Electronic structure of \GTS{} without SOC. (a) Cubic structure (F$\bar{4}$3m, a=10.3849~\AA). (b) Tetragonal structure (P$\bar{4}$2$_{1}$m, a=10.4677~\AA, c=21.0086~\AA). In tetragonal structure the electronic density of states is reduced at the Fermi level and opens a small gap below \Ts{}.}
\label{fig:SIBandStruct}
\end{figure}

\begin{figure}[t!]
 \includegraphics{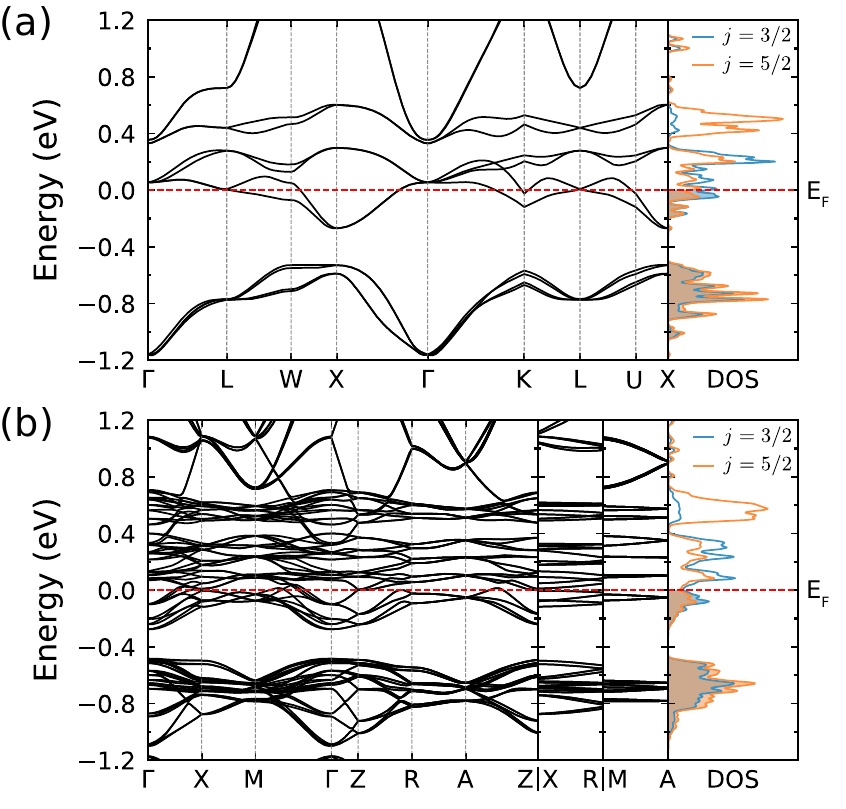}
\caption{Electronic structure of \GTS{} with SOC, with the density of states projected into atomic j=3/2 and j=5/2 basis shown in the side panel. (a) Cubic structure (F$\bar{4}$3m, a=10.3260~\AA). (b) Tetragonal structure (P$\bar{4}$2$_{1}$m, a=10.3988~\AA, c=20.8129~\AA). Both structures require electronic correlation (U) to open a gap, therefore, retains a Mott insulator below \Ts{}.}
\label{fig:SIBandStructSOC}
\end{figure}

The calculated band structures and electronic density of states for both PAW-PBE and PBE+SOC calculations are shown in Fig.~\ref{fig:SIBandStruct} and Fig.~\ref{fig:SIBandStructSOC}. In both calculations, the density of electronic states at the E$_{F}$ is drastically reduced from that of the cubic cell. Without spin-orbit coupling, the unit cell doubled structure opens a small gap since the total electron number is even in a primitive cell. However, with spin-orbit coupling, the splitting of the electronic bands retains a metallic state in the unit cell doubled structure. A large Coulomb interaction, U, is required to open an insulating gap and our band structure calculations indicate that \GTS{} is retains its Mott insulating nature at low tempertures.

\subsection{Phonon calculations}
We used \textsf{Phonopy} package \cite{phonopy} for ab-initio phonon calculations and \textsf{Quantum ESPRESSO} \cite{giannozzi2009quantum,giannozzi2017advanced} as the force calculator. For phonon calculation, we used 680 eV cutoff for plane-wave energy with a 2$\times$2$\times$1 Monkhorst-Pack k-point mesh for self-consistent-field calculations. We checked the total energy convergence with the cutoff energy and k-point mesh for calculations. The minimal k-point were used for the cell with 104 atoms. To compare with the results measured by inelastic neutron scattering, the neutron weighted phonon DOS was calculated by neutron cross-section for each atom and the projected phonon density of states as follows
\begin{equation}
PDOS(E) = \sum_{i} \sigma_{i}g_{i}(E),
\end{equation}
where $\sigma_{i}$ and $g_{i}(E)$ are neutron cross-section and projected phonon density of states for each species of atom. To simulate the measured PDOS, we convolved the calculated PDOS with an empirical instrumental resolution function for SEQUOIA  \cite{granroth2010sequoia}
\begin{multline*}
f(E) = 1.435 \times 10 ^{-6} \times E ^{3} +
3.885 \times 10 ^{-4} \times E ^{2} \\ - 7.882 \times 10 ^{-2} \times E + 4.295.
\end{multline*}

The irreducible representation of phonon modes can be calculated with the information of occupied Wyckoff positions (WPs). In P$\bar{4}$2$_{1}$m tetragonal structure, Ga atoms occupy 4e WPs. Ta and Se atoms are on 4e and 8f WPs. Each atom on 4e site has 2A$_{1}$ + A$_{2}$ + B$_{1}$ + 2B$_{2}$ + 3E optical phonon modes, and 3A$_{1}$ + 3A$_{2}$ + 3B$_{1}$ + 3B$_{2}$ + 6E for atoms on 8f WPs. Therefore,  the entire unit cell with 20 asymmetric atoms has 46A$_{1}$ + 32A$_{2}$ + 32B$_{1}$ + 46B$_{2}$ + 78E optical modes. As Fig.~\ref{fig:4} in main text shows, the most distinguishable difference between cubic and tetragonal structure is the modes around 13 meV. From our DFT calculation, we assigned phonon modes between 12 and 14 meV to irreducible representation 2A$_{1}$ + 2A$_{2}$ + 2B$_{1}$ + 3B$_{2}$ + 4E, which is the mixed modes of all Ga, Ta and Se atoms.


%

\end{document}